\documentstyle[aps]{revtex}

\makeatletter


\begin{document}

\author{D. Reguera, J.M. Rub\'{\i}, A. P\'{e}rez-Madrid}

\title{Controlling anomalous stresses in soft field-responsive systems.}

\address{Departament de F\'{\i}sica Fonamental-CER on Physics of Complex Systems\\ Facultat de F\'{\i}sica,Universitat de Barcelona, Diagonal 647, 08028 Barcelona, Spain}

\maketitle
\begin{abstract}
We report a new phenomenon occurring in field-responsive suspensions: shear-induced
anomalous stresses. Competition between a rotating field and a shear flow originates
a multiplicity of anomalous stress behaviors in suspensions of bounded dimers
constituted by induced dipoles. The great variety of stress regimes includes
non-monotonous behaviors, multi-resonances, negative viscosity effect and blockades.
The reversibility of the transitions between the different regimes and the self-similarity
of the stresses make this phenomenon controllable and therefore applicable to
modify macroscopic properties of soft condensed matter phases. 
\end{abstract}

\section{introduction}

Field-responsive systems constitute a class of soft condensed matter systems
undergoing significant responses leading to important macroscopic changes upon
application of an external field \cite{ferrohyd}-\cite{grana}. This peculiar
characteristics has been used in many applications and may become useful in
the implementation of different devices \cite{bull}. Electro- and magneto-rheological
fluids, ferrofluids and magnetic holes are typical examples of field-responsive
systems which have been subject of many recent investigations\cite{cadenas}-\cite{kn:bossis}.

These systems consist essentially in two phases, one is a dispersion of \emph{smart}
active units, whereas the other is a liquid, or more generally a soft phase,
practically inactive to the action of the field. 
The mechanical response of such units to the applied field depends on their
nature. If the particles bear permanent dipoles, they induce stresses in the
liquid phase during their reorientation process even in the single-particle
domain. When the dipoles are induced their dipolar moments are always collinear
with the field. Therefore in this case the only way to induce mechanical responses
is through the formation of assemblies of particles, which occurs at higher
concentrations, when dipolar interactions start to play a significant role.
The elementary assembled unit exhibiting mechanical response is a bounded pair
of induced dipoles (dimer).

Our purpose in this paper is to show that stresses induced by these field-responsive
elementary units, the bounded dimers, exhibit a multiplicity of regimes emerging
from the nonlinear nature of the dynamics, not observed in other field-responsive
phases analyzed up to now. The stresses are anomalous as they do not necessarily
vary monotonously with the characteristic parameters and reversible as their
appearance is not subjected to intrinsic structural changes of the system. This
peculiar property has an important consequence: it can be used to control the
induction of stresses in the solvent phase.

We have organized the paper in the following way. In Section II we introduce the model describing the dynamics of the system. Section III is devoted to analyze the stresses generated by the particles, whereas in section IV we discuss the rheology of the suspension. Finally, the last section is intended as a summary of our main results.

\section{The model}

To illustrate this phenomenon, we consider a 2D model in which the dynamics
of the orientation \( \varphi  \) of the bounded dimer captures the two basic
ingredients present in experimental situations, namely a pure rotation caused
by the applied field and a term breaking the symmetry of the dynamics which
originates from the presence of a shear flow:

\begin{equation}
\label{ec_dina}
\dot{\varphi }=-A(t)\sin (2(\varphi -\alpha (t)))-b\sin ^{2}\varphi 
\end{equation}
 where the overdot denotes total time derivative. Here we have considered the
general case in which the pure rotation is modulated by the frequency \( w_{h} \)
by 
\begin{equation}
\label{a}
A(t)=w_{c}(cos^{2}(w_{h}t)+r^{2}sin^{2}(w_{h}t))
\end{equation}
 where \( w_{c} \) is a characteristic frequency and \( r \) denotes its degree
of polarization ranging from \( r=1 \) corresponding to circular polarization
to \( r=0 \) holding for linear polarization; \( \alpha (t) \) is a time dependent
phase and \( b \) is the shear rate.

Physical realizations of this model are in general the 2D dynamics of a bounded
pair of spherical induced dipoles in the presence of a shear flow with velocity
profile \( b \)\( y\bf {\widehat{x}} \), and of an external rotating field
with frequency \( \omega _{h} \) and components \( H_{x} \)\( \cos \left( \omega _{h}t\right) \bf {\widehat{x}} \)
and \( H_{y} \)\( \sin \left( \omega _{h}t\right) \bf {\widehat{y}} \). The
equation of motion of the rotating dimer Eq. \ref{ec_dina} then emerges from
balancing out the hydrodynamic torque 
\begin{equation}
\label{torque h}
T_{hy}=-6\pi \eta _{0}(b\sin ^{2}\varphi +\dot{\varphi })
\end{equation}
 and the external field torque 
\begin{equation}
\label{torque m}
T_{M}=-6\pi \eta _{0}A(t)\sin (2(\alpha (t)-\varphi ))
\end{equation}
 arising from the energy of dipolar interaction 
\begin{equation}
\label{potencial}
U(\varphi )=\frac{M_{V}^{2}(1-3\cos ^{2}(\alpha (t)-\varphi ))}{d^{3}}
\end{equation}
In the previous equations, \( \eta _{0} \) is the viscosity of the liquid phase,
\( \alpha (t)=\arctan (r\tan (w_{h}t)) \) is the direction of the field, and
\( M_{V}=V\chi _{eff}H \) is the induced moment with \( \chi _{eff} \) the
effective susceptibility and \( V \) is the volume of the sphere with diameter
\( d \). Within this context, the value of the characteristic frequency can
be identified with \( w_{c}=\frac{\chi _{eff}^{2}H^{2}_{x}V^{2}}{2\pi \eta _{0}d^{3}} \),
and \( r=H_{y}/H_{x} \). The model, for a magnetic rotor in the absence of
the symmetry-breaking term has been discussed by Skjeltorp {\em et al.} in the context of
the nonlinear dynamics of a bounded pair of magnetic holes \cite{kn:skjeltorp}-\cite{kn:skjeltorp1}.
For the particular case \( w_{c}=\frac{mH_{x}}{6\pi \eta _{0}} \)and \( b=0 \)
Eq. \ref{ec_dina} also describes the 2D dynamics of a ferrofluid particle
with magnetic moment \emph{\( m \) in} a static magnetic field and a vorticity
field \( -w_{h}\bf {\widehat{z}} \), in the absence of noise. \cite{kn:agusti}

The motion of the pair induces stresses in the whole system emerging from the
conversion of field torque into tensions in the fluid. In this sense, this process
can be viewed as a mechanism of transduction of field energy into stresses whose
efficiency is determined by the dynamics. The induced stress is simply the averaged
density of hydrodynamic torque:

\begin{equation}
\label{stress}
S=6\eta _{0}c\langle \dot{\varphi }+b\sin ^{2}\varphi \rangle 
\end{equation}
 where \( c \) represents the volume fraction of dimers.

\section{Dynamics and stresses}

\subsection{Circular polarization}

In order to elucidate the main features of this model, we have solved numerically
Eq. \ref{ec_dina}. In Figure 1a we have depicted the stress as a function of
the frequency of the field, for different values of shear rate corresponding
to the case of circular polarization (\( r=1 \)).Since the stress is an homogeneous
function, \( S(\lambda w_{c},\lambda b,\lambda w_{h})=\lambda S(w_{c},b,w_{h}) \),
its behavior can be analyzed in terms of the scaled quantities \( b^{\prime }=b/w_{c} \),
\( w_{h}^{\prime }=w_{h}/w_{c} \) and \( S^{\prime }(w^{\prime }_{h},b^{\prime })=S/w_{c} \).

In the absence of shear flow (\( b'=0 \)), the interplay between hydrodynamic
and field effects originates two basic dynamic regimes determined by the value
of \( \left| w_{h}'\right|  \). When this frequency is smaller than the threshold
\( \left| w^{\prime }_{h}\right| =1 \), the dimer follows the field with a
fixed phase-lag and the same angular velocity, performing {\em uniform} oscillations. At frequencies higher than the
threshold value the system is not longer able to follow the field and undergoes
periodic rotations with stops and backwards motions (''jerky'' oscillations)
\cite{kn:skjeltorp}. These two modes of motion are manifested in two different
regimes of the stress in Fig 1a. A linear regime, for \( \left| w^{\prime }_{h}\right| <1 \),
in which the scaled stress is just the frequency of the field and a monotonous
decay regime for \( \left| w_{h}'\right| >1 \), where the modulus of the stress
decrease due to jerky oscillations. During backwards rotations field energy
is wasted inducing ``wrong sign tensions''. When they become as important
as forward rotations, which occurs at high frequencies, the net transduction
of energy, and consequently the induced stress, is practically inexistent.

The presence of a shear flow completely modifies the dynamical response leading
to the appearance of a richer phenomenology. The role played by the flow is
manifold. On one hand, it breaks the symmetry of the dynamics by fixing a direction
of rotation which implies that the property \( S'(w_{h}')=-S'(-w_{h}') \),
which holds in the absence of shear, is no longer valid. On the other hand,
the regimes in which one of the two competing rotational mechanisms, related
to the field and to the flow, dominates are intrinsically different. Finally,
the presence of the new time scale \( b \) is responsible for the appearance
of new \emph{synchronization} mechanisms.

For \( b'<1 \) the strength of the field dominates and the behavior of the
stress is similar to that of the case \( b'=0 \) but shifted in frequency by
an amount \( b'/2 \) as one can notice in Fig 1a. In Fig 1b, we have represented
some snapshots of the dynamical modes of rotation corresponding to the different
regimes in the stress. Upon increasing \( w_{h}' \) we can generate the sequence
of modes: jerky(first two snapshots)-uniform-jerky-localized oscillations. We
have found a dynamic transition, from \emph{jerky} to \emph{localized oscillations,}
at a characteristic positive frequency, which depends on \( b' \), with its
subsequent macroscopic consequences in the stress. Moreover, competition between
flow and field involves the breaking of the symmetry of the stress and leads
to the decrease of the stress peak at positive frequencies. 

In the opposite case, when \( b'>1 \), the effects of the flow dominate and
in this situation, even at frequencies near to zero, the rotation imposed by
the field is very different from the one dictated by the shear. As we can see
in Fig. 1a, the positive peak has definitively disappeared and the behavior
of the stress is characterized by the development of small \emph{multi-resonances}
followed by linear increases and decreases of the stress with slopes -1, -2...
The origin of this behavior is the synchronization of the field and the shear,
exciting mode-locks of the pair with frequencies ratio \( S':w_{h}' \) 1:1
or 2:1, etc.   It is important to highlight that, for every value of the shear rate, stress
curves overlap at high and moderate frequencies. 

As an illustrative
example, in Fig 1c we have depicted some snapshots of the dynamics for \( b'=8 \)
corresponding to the different regimes of the stress obtaining a sequence of
modes: jerky-uniform-jerky (3rd and 4th snapshots)-localized oscillations, upon
increasing \( w_{h}' \).

It is worth pointing out that the frequency of the negative stress minimum, where transition between linear
and jerky oscillations regime occurs, follows a power law in terms of the shear
rate: \( w_{\min }^{\prime }\sim -(b^{\prime }+1)^{0.45} \), as it is shown in Fig. 2.

\subsection{Elliptical polarization}

Even more interesting is the case of elliptical polarization. In Fig 3a we have
depicted the stress against frequency for \( r=0.5 \) and different values
of the shear rate. In the absence of shear, the dynamics basically exhibits
three different modes as we increase \( \left| w_{h}'\right|  \): i) a \emph{phase-locked}
mode, where the system performs modulated (by the term \( A(t) \)) uniform
rotations with average frequency \( w_{h}' \); ii) a \emph{modulated ''jerky''} oscillations mode above a critical frequency
; iii) and \emph{localized oscillations}, with null average velocity above another
characteristic frequency.These modes
are responsible for three different behaviors in the stress observed in Fig
2a: a linear regime near \( \left| w_{h}'\right| =0 \) , a decay in the modulus
due to jerky oscillations and a non-stress zone when the net rotation vanishes,
respectively.

The introduction of the flow changes these regimes significantly. For very low
values of \( b' \), the modes of rotation are slightly modified, as shown in
Fig 3b; the curve is simply shifted by \( b^{\prime }/2 \) in frequencies;
and the scaled stress when localization appears is no longer zero but saturates
at approximately \( b' \), positive even for \( w_{h}'<0 \).

At a critical value of the shear rate, the positive stress maximum disappears
as shown in Fig 3a for \( b'=0.5 \). Above this value of the shear rate, \emph{multi-resonances}
develop near \( w_{h}'\sim 0 \). The critical frequency denoting the transition
from uniform to jerky oscillations, corresponding to the position of the minimum
of the stress, is shifted following a power law with an exponent near 0.5. Additionally,
jerky oscillation mode for negative frequencies persist in a wider range, which
causes in turn persistence of the negative stress region at moderate/ high frequencies.
The dynamic transition from jerky to localized oscillations at negative frequencies
is the signature of a change in the sign of the stress.

\section{Rheology}

When represented as a function of the shear rate the stress exhibits a wide
variety of different anomalous behaviors. This feature contrasts with the monotonous
behavior observed in systems inert to the applied field \cite{kn:larson}. The
existence of such a rich phenomenology is manifested in Figs. 4 and 5. Their
most salient feature is that upon fixing a proper value of the frequency of
the field, we can monitor and promote drastic changes in the mechanical response
of the system. 

For some determined values of the shear rate the induced stress
has a steep increase. Consequently, the system exhibits a \emph{multi-resonant
response}, as we can see for \( w_{h}'=0.5 \) in Fig 4. These resonances originate
from the synchronization of the field with the hydrodynamic response of the
system, which enhances the induction of tensions. 

For a wide range of values
of the frequency and a wide interval of values of the shear rate the response
of the system to the variations of the shear rate is inhibited (\emph{no-response}
or \emph{``blockade'' regime}, corresponding to the flat curves in Fig. 4).

There also exists a regime where the transducted field energy improves the rotation
of the pair in the shear, leading to a reduction of the apparent viscosity of
the fluid. This phenomenon \cite{kn:agusti}-\cite{kn:bacri} has been referred
to as the \emph{negative viscosity effect.} 

 Finally, there
appears monotonous \emph{shear thickening}, \emph{shear thinning} regimes or
combinations of both as is manifested in Fig 4. 

It is worth pointing out that the stress curves are quite self-similar, as Fig 6 manifests. 
Therefore, we can tune the regime we are interested in by properly modifying
one of the parameters of the problem. Moreover, the existence of the scaling
invariance ensures its accessibility in all the range of values.

\section{Discussion and conclusions}

In this paper, we have shown the possibility of generating stresses of very different
nature in assemblies of pairs of induced dipoles. The implementation of a model
which mimics the dynamics of the field-responding unit leads to the appearance
of a rich variety of nonlinear stress regimes involving multi-resonances, shear-
thickening and thinning, negative viscosity or blockades.

This multiplicity of intrinsically different behaviors together with the reversible
nature of the transition mechanisms can be utilized to control the induction
of stresses in the inactive phase. A broad field of applications of this phenomenon
can then be open. The importance of the control of the stress lies in the fact
that stress itself may induce significant modifications in soft condensed matter
phases. To mention just a few examples, stresses may induce structural transitions
in surfactant solutions\cite{yama} or gelation\cite{pine}; they can also modify
the orientation of surfactant phases, liquid-crystals\cite{safinya} or polymers\cite{science}.
Moreover, alterations in the distribution of stresses may lead to important
changes in the rheological properties of the system\cite{kn:rosensweig}.

In the cases we have analyzed, possible noise sources have not been considered. Whereas absence of noise constitutes a good approximation for large particles, as magnetic holes\cite{kn:skjeltorp}, smaller particles, as in the case of a ferrofluid, are affected by Brownian torques. In the first case, the model we have proposed through Eq. \ref{ec_dina} is enough to describe the dynamics of the suspended phase. For the ferrofluid, however, the model must and can be easily generalized to include noise sources. 

 Our findings may open
new perspectives for research in these systems offering some insight into the
mesoscopic mechanisms controlling macroscopic nonlinear behaviors.

\section{Acknowledgments}

We would like to thank T. Alarc\'{o}n for valuable discussions. This work has
been supported by DGICYT of the Spanish Government under grant PB98-1258, and
by the INCO-COPERNICUS program of the European Commission under Contract IC15-CT96-0719.
D. Reguera wishes to thank Generalitat de Catalunya for financial support.


\begin{figure}

\caption{a) Scaled stress as a function of the frequency of the field for some representative
values of \protect\( b'\protect \) for the case of circular polarization. Below
we include polar plots of \protect\( \varphi \protect \) vs. time (radially)
representing the dynamical regimes for different values of \protect\( w_{h}'\protect \)
and for values of the scaled shear rate b) \protect\( b'=0.5\protect \) c)
\protect\( b'=8.0\protect \)}
\end{figure}

 \begin{figure}

\caption{Scaling law for the frequency corresponding to the negative stress minimum for different values of the shear rate, in the case of circular polarization.}
\end{figure}

\begin{figure}
\caption{a) Scaled stress as a function of the frequency of the field for some representative
values of \protect\( b'\protect \) for the case of r=0.5. Below we include
polar plots of \protect\( \varphi \protect \) vs. time (radially) representing
the dynamical regimes for different values of \protect\( w_{h}'\protect \)
and for values of the scaled shear rate b) \protect\( b'=0.2\protect \) c)
\protect\( b'=8.0\protect \)}
\end{figure}

\begin{figure}
\caption{Scaled stress \protect\( S'\protect \) as a function of the shear rate \protect\( b'\protect \)
for some values of the frequency of the field \protect\( w_{h}'\protect \)
as well positive (against the rotation induced by the flow) as negative (in
the same direction as the rotation of the flow) for the case of circular polarization.}
\end{figure}

\begin{figure}
\caption{Scaled stress \protect\( S'\protect \) as a function of the shear rate \protect\( b'\protect \)
for some values of the frequency of the field \protect\( w_{h}'\protect \)
as well positive (against the rotation induced by the flow) as negative (in
the same direction as the rotation of the flow) corresponding to \protect\( r=0.5\protect \).}
\end{figure}

\begin{figure}
\caption{Scaled stress for some values of the shear rate, as a function of the scaled frequency $w'_{h} \sqrt{b'_{0}/b'}$. We have represented the case $b'_{0}=32$, where the self-similarity of the curves becomes manifested.}
\end{figure}

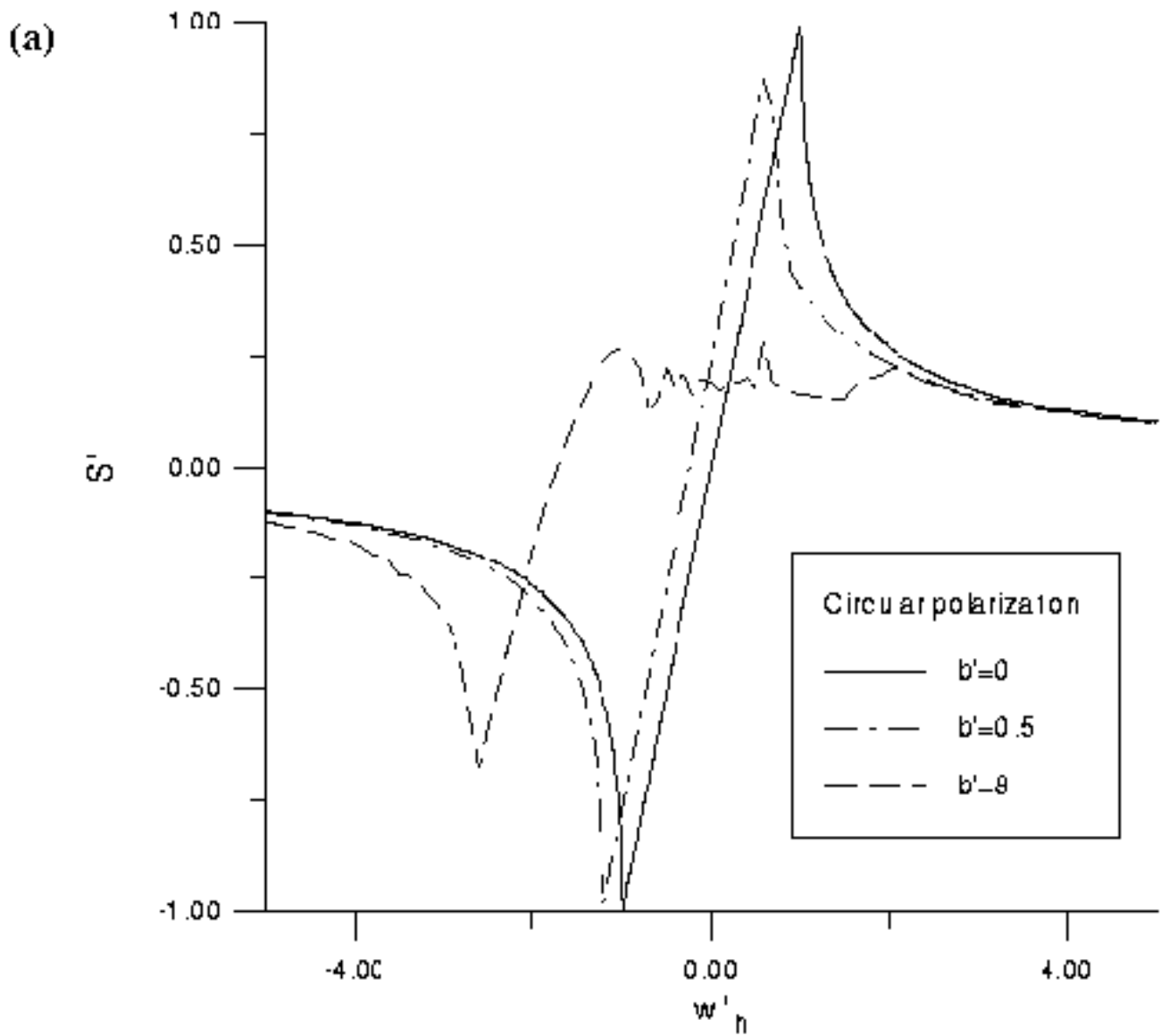
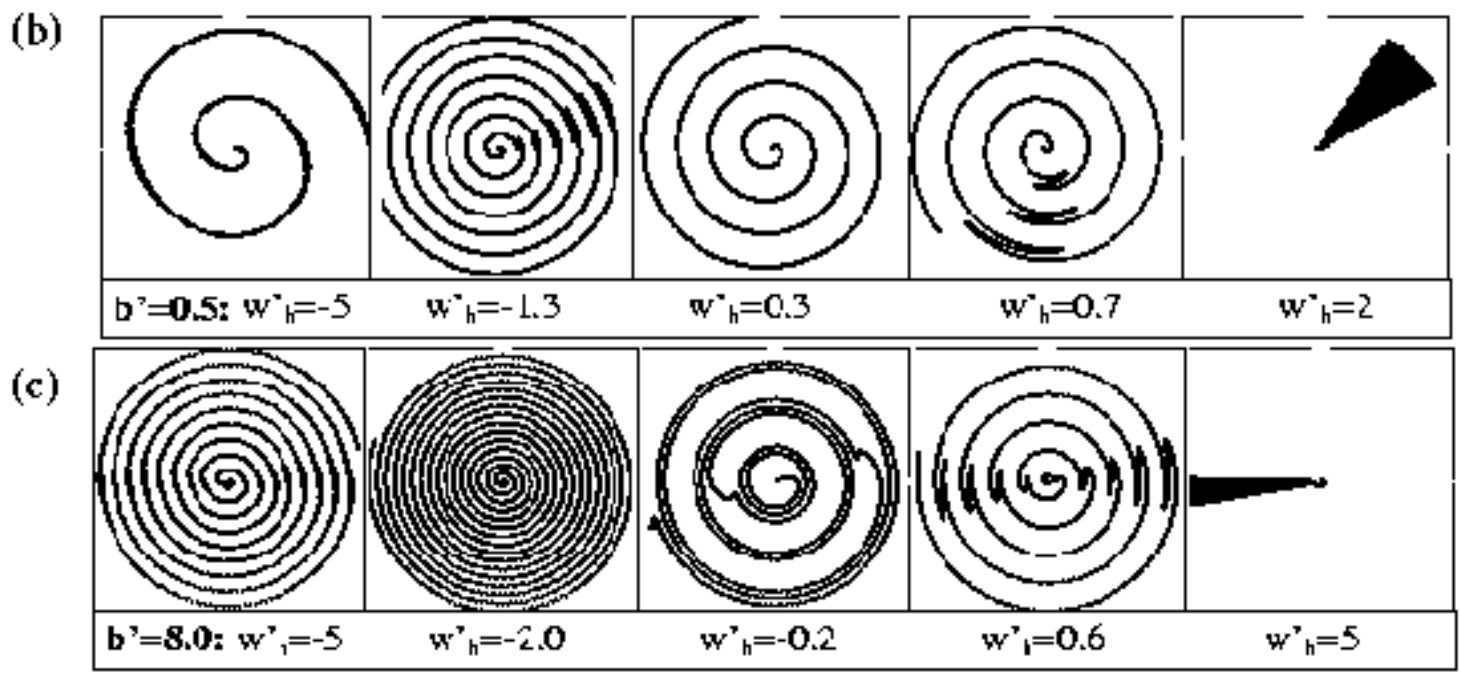
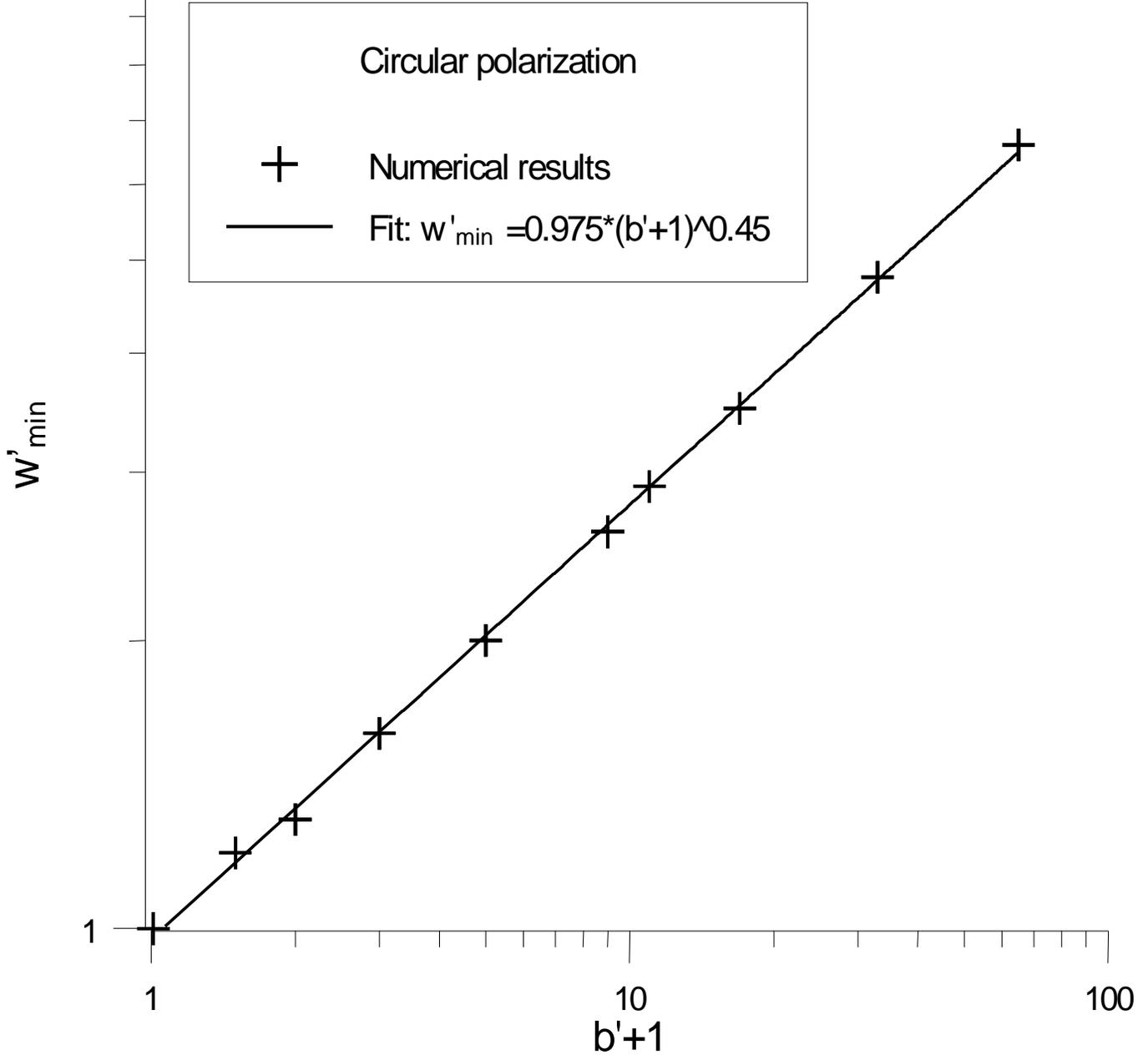
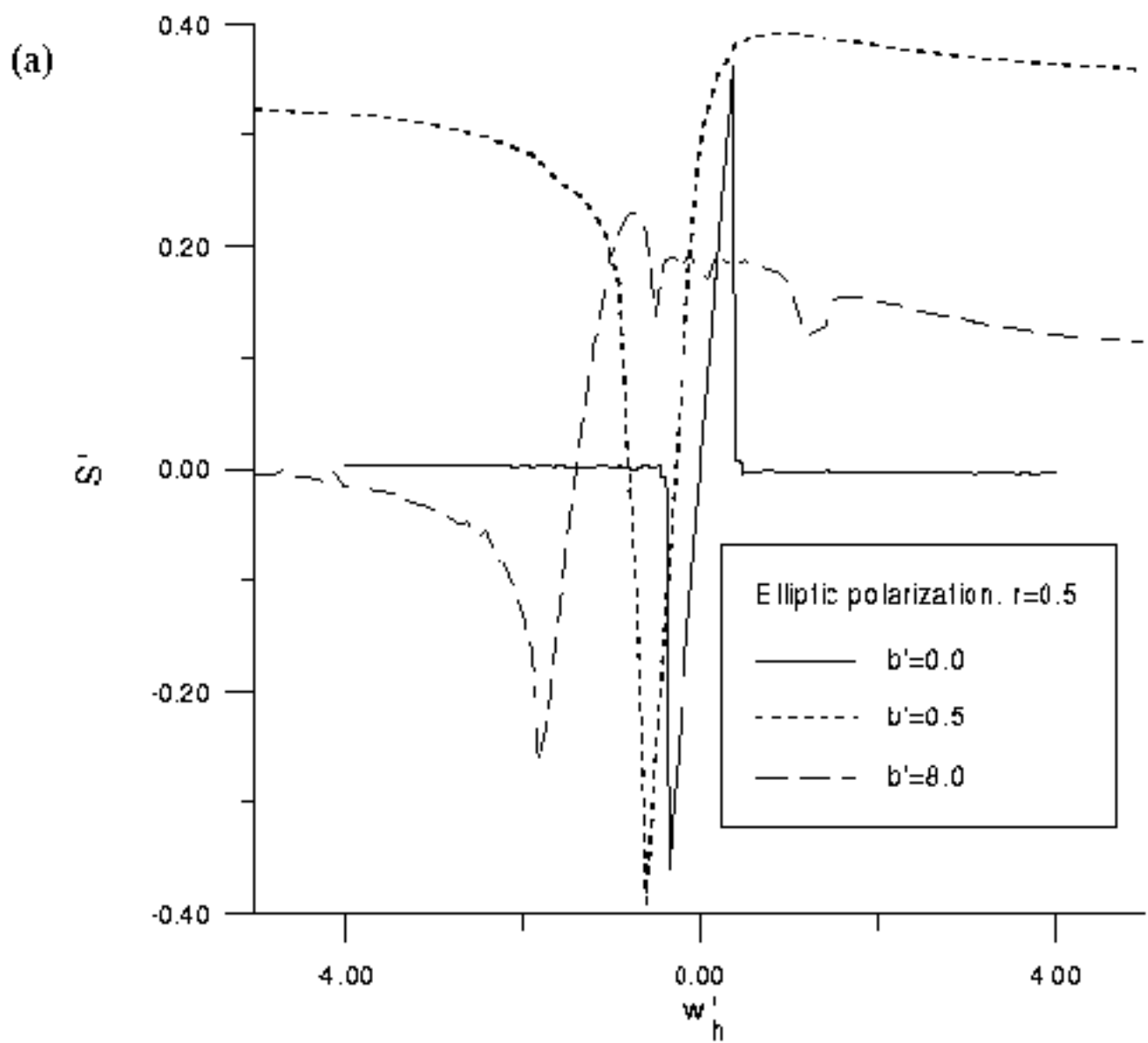
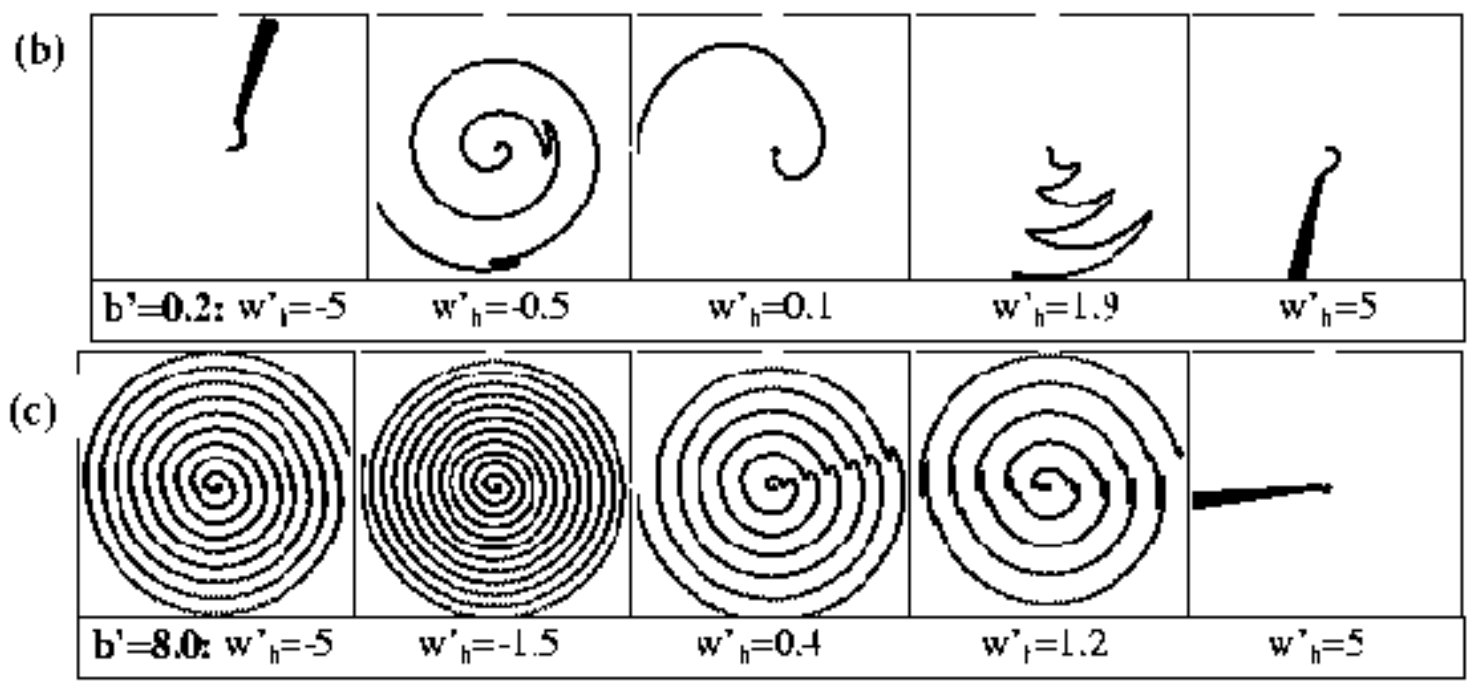
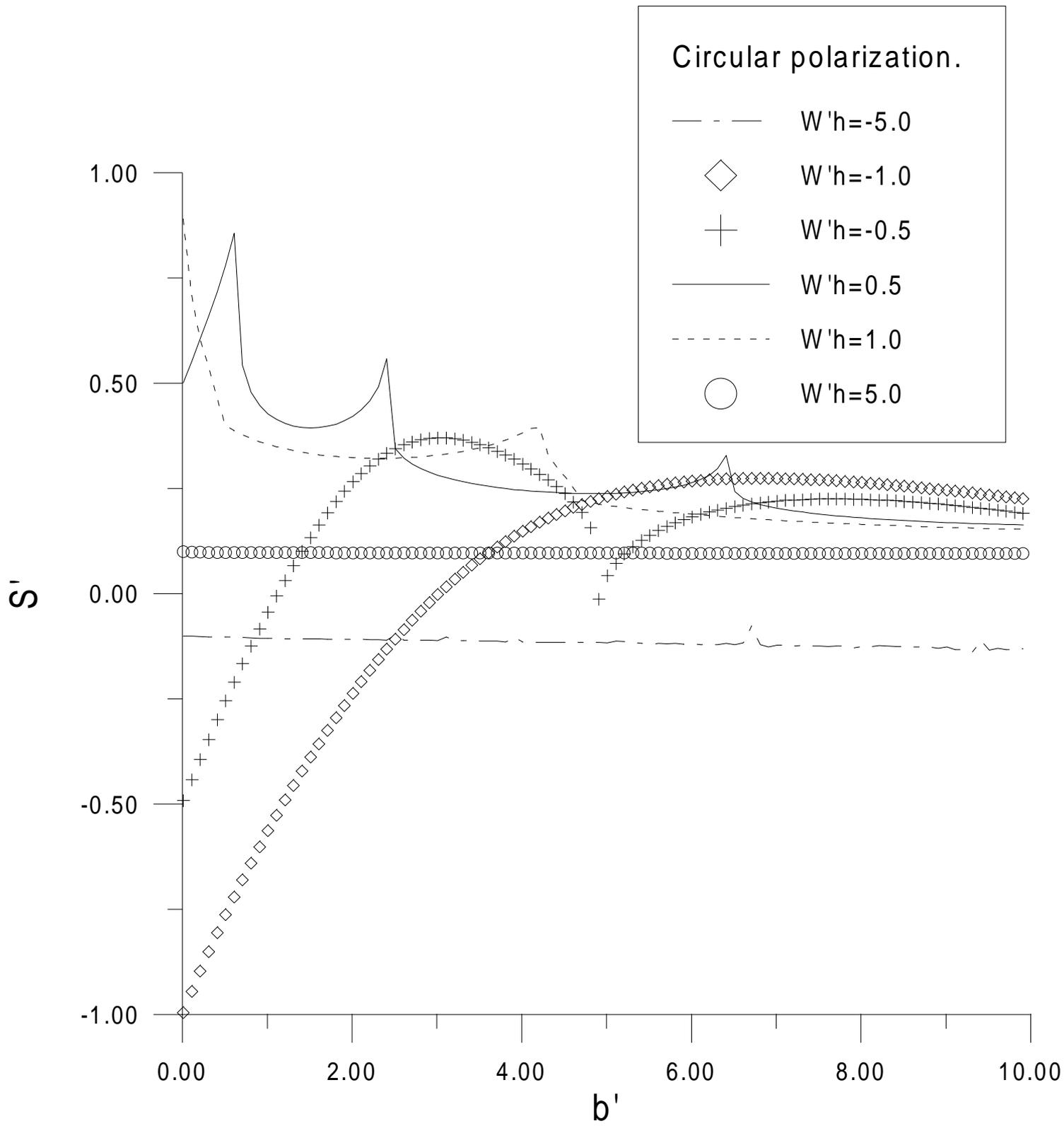
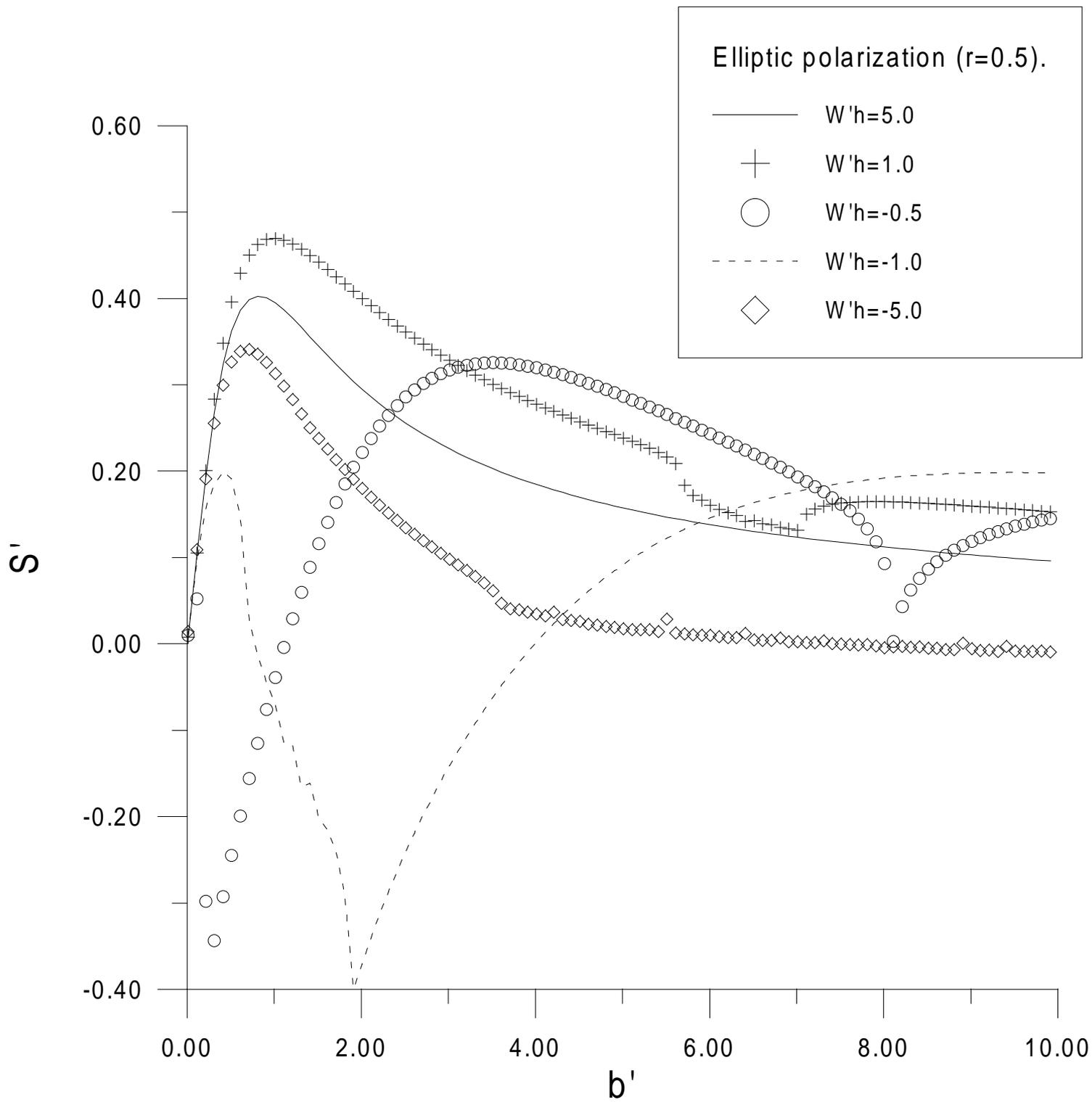
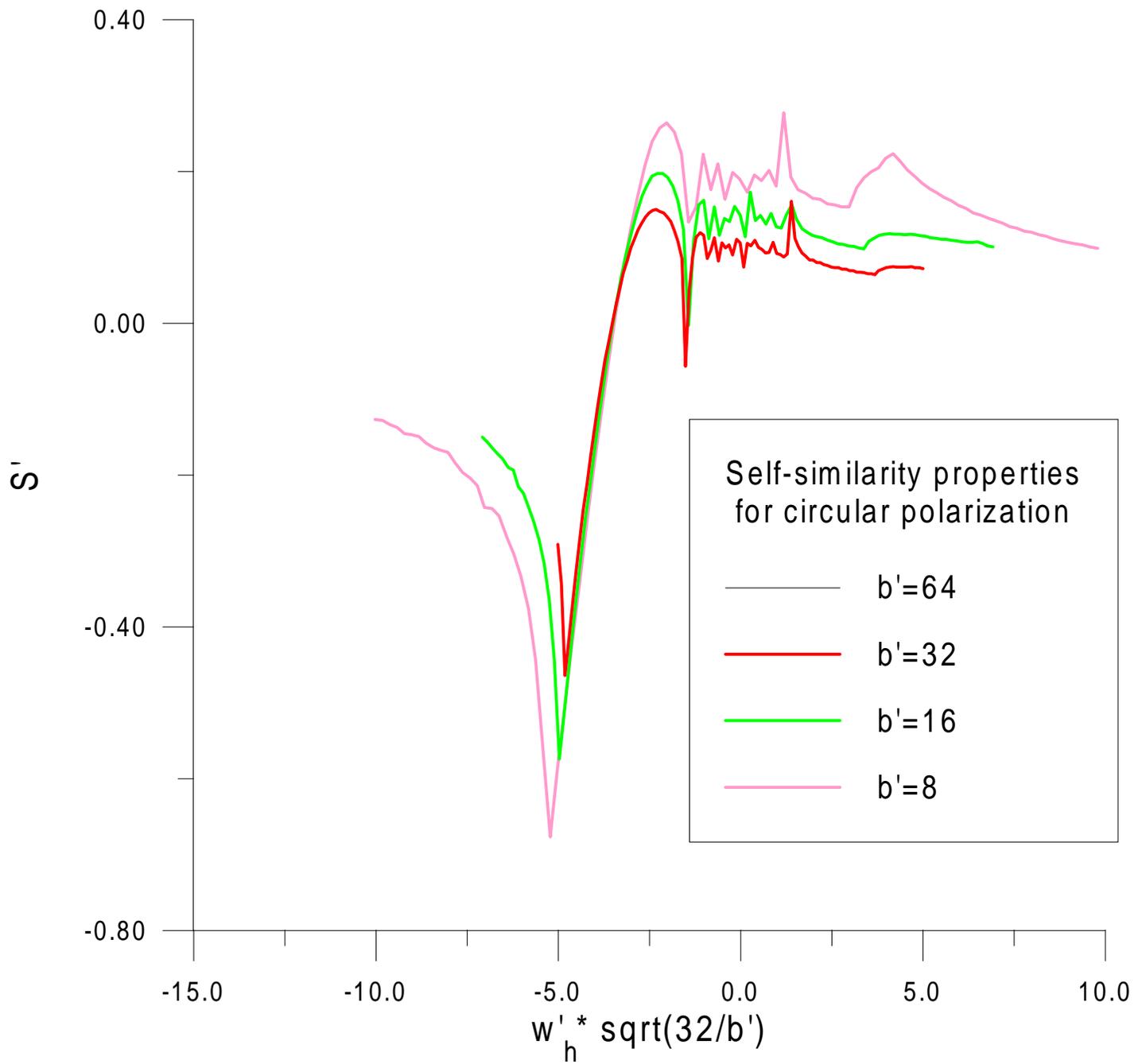
\end{document}